\documentclass[11pt]{article}
\usepackage[dvipdfmx]{graphicx}
\usepackage{setspace}
\usepackage{latexsym}
\usepackage{amssymb}
\usepackage{color}
\usepackage{amsfonts}
\usepackage{amsmath}
\usepackage{booktabs}
\usepackage{caption}
\usepackage{subcaption}
\usepackage{longtable}
\usepackage{mathtools}
\usepackage{multicol}
\usepackage{graphicx} 
\usepackage{wrapfig} %
\usepackage{float}
\usepackage[whole]{bxcjkjatype}
\usepackage{indentfirst}

\newcommand{\BE}{\begin{equation}}
\newcommand{\EE}{\end{equation}}
\newcommand{\BEQ}{\begin{eqnarray}}
\newcommand{\EEQ}{\end{eqnarray}}
\newcommand{\BEQA}{\begin{eqnarray*}}
	\newcommand{\EEQA}{\end{eqnarray*}}
\newcommand{\BA}{\begin{array}}
	\newcommand{\EA}{\end{array}}

\begin{document}

	\begin{spacing}{2.0}
	\baselineskip= 7mm
	\begin{center}
		{\Large\bf Improving volatility forecasts of the Nikkei 225 stock index using a realized EGARCH model with realized and realized range-based volatilities}
		\vspace{15mm}\\
	 Yaming Chang\vspace{5mm}\\

\begin{abstract}
		\textbf{
This paper applies the realized exponential generalized autoregressive conditional heteroskedasticity (REGARCH) model to analyze the Nikkei 225 index from 2010 to 2017, utilizing realized variance (RV) and realized range-based volatility (RRV) as high-frequency measures of volatility. The findings show that REGARCH models outperform standard GARCH family models in both in-sample fitting and out-of-sample forecasting, driven by the dynamic information embedded in high-frequency realized measures. Incorporating multiple realized measures within a joint REGARCH framework further enhances model performance. Notably, RRV demonstrates superior predictive power compared to RV, as evidenced by improvements in forecast accuracy metrics. Moreover, the forecasting results remain robust under both rolling-window and recursive evaluation schemes. }
\end{abstract}
\end{center}
\vspace{5mm}		
\end{spacing}
\newpage
	\begin{spacing}{2.0}
\section{Introduction}


In finance, volatility measures the dispersion of returns for a given security or market index. It is commonly estimated using the standard deviation or variance of returns. Accurate volatility forecasting plays a crucial role in investment decisions, security valuation, risk management, and monetary policy. It is also a key parameter in pricing financial derivatives. As a result, studying the forecasting performance of various volatility models has remained a focal point for both academics and practitioners.

Researchers have devoted significant attention to modeling daily and high-frequency volatility processes, as reflected in the growing body of empirical literature.  Many studies have employed the autoregressive conditional heteroskedasticity (ARCH) family models including Engle's ARCH model (1982)\nocite{ARCH}, Bollerslev's Generalized ARCH (GRACH) model (1986)\nocite{GARCH}, and the stochastic volatility (SV) models (Taylor, 1986)\nocite{SV}. These models have been further extended to capture asymmetric volatility dynamics in financial markets, leading to the development of Nelson's Exponential GARCH (EGARCH) (1991)\nocite{EGARCH}, Glosten et al.'s GJR (1993)\nocite{GJR}, Ding et al.'s asymmetric power GARCH (1993)\nocite{APGARCH}, and Taylor's asymmetric SV (1994)\nocite{ASV}).   

 Another related strand of research focuses on high-frequency realized measures of volatility. Andersen and Bollerslev (1998)\nocite{Andersen1998} demonstrated that ex-post daily foreign exchange volatility is best measured by aggregating squared five-minute returns. While the high-frequency data-based volatility measures offer richer dynamic information than those based on daily data, they are subject to challenges such as non-trading hours and microstructure noise. 
 
 The absence of trading during certain hours can lead to an underestimation of true volatility when calculating 24-hour volatility. Additionally, microstructure noise—arising from bid-ask bounce, nonsynchronous trading, infrequent trading, and price discreteness—can introduce biases in realized measures, as discussed by Campbell et al., (1997)\nocite{Micro-noise}, Madhavan (2000)\nocite{Madhavan}, and Biais et al. (2005)\nocite{Biais}. Various methods have been proposed to mitigate microstructure noise in realized volatility estimation (Zhang et al., 2005\nocite{Zhang}; Zhang, 2006\nocite{Zhang2006}; Bandi and Russell, 2008, 2011\nocite{Bandi2008}\nocite{Bandi2011}; Barndorff-Nielsen et al., 2008, 2011\nocite{RK2008}\nocite{RK2011}). Recent research by Liu et al. (2015)\nocite{5min} suggests that five-minute realized variance (RV) remains a robust benchmark, with limited scope for significant improvements. Accordingly, this study calculates realized measures using a five-minute frequency. To further evaluate the effectiveness of realized measures, we incorporate realized range-based volatility (RRV)\nocite{RRV}, introduced by Christensen and Podolskij (2007), as a potentially more efficient alternative to the dominant five-minute RV.

A considerable amount of recent literature has extended conventional volatility models by including high-frequency data-based realized measures to improve estimation goodness and model forecast accuracy. Takahashi et al. (2009)\nocite{RSV} have extended the SV model, while Hansen et al. (2012) \nocite{RGARCH} proposed the extension of GARCH models, incorporating them with realized volatility. Hansen et al. (2016) recently applied \nocite{JREGARCH} extended realized EGARCH model that can use multiple realized volatility measures for the modeling of a return series when forecasting volatility. These models can also adjust the realized volatility bias caused by microstructure noise and non-trading hours.

This article proposes the forecasting of conditional volatility by applying a recently suggested realized EGARCH model to the Nikkei 225 stock index. We shall also attempt to extract volatility information from multiple realized measures, the RV and  RRV, and to incorporate these measures into a volatility model to enhance the forecast quality.  The advantages of using a joint realized EGARCH model with multiple volatility measures in volatility forecast are straightforward. First, the  model augments the conventional EGARCH model with the dynamic information of realized measures estimating the joint dependence between the ex-post realized measures and the ex-ante conditional variance. Second, by regressing the realized measure on the conditional variance of returns, the realized EGARCH model can adjust the realized volatility bias caused by microstructure noise and non-trading hours. While realized GARCH-type models have already been applied to quantile forecasts by Watanabe (2012) and to option pricing by Huang et al. (2016), this article presents the first-known research applying a joint realized EGARCH model, including intraday range-based volatility, to volatility forecasts in the stock market. The benefits of using range-based volatility measures have been demonstrated in  Alizadeh et al. (2002)\nocite{dailyrange}, Brandt and Jones (2006)\nocite{Brandt}, and Christensen and Podolskij (2007, 2009)\nocite{RRV}\nocite{BCRRV}.

 The research in this article is closely related to Hansen et al. (2016)\nocite{JREGARCH}, while offering significant differences from that study. Firstly, the forecast evaluation in Hansen et al. (2016) was measured using only one criterion, out-of-sample log-likelihood, yet this research retains the old criterion and employs robust forecast comparison methods suggested by Patton (2011)\nocite{Patton_fore}.  Secondly, our evidence is compelling under both recursive and rolling-window schemes. Thirdly, in our empirical application, the use of an intraday range as the realized measure in the realized EGARCH models increases the out-of-sample forecasting ability more than the use of RV. Ultimately, our results indicate that the joint realized EGARCH model, which incorporates both RV and RRV, outperforms in forecast accuracy.

The remainder of this paper is organized as follows: Section 2 introduces the conventional GARCH family models, realized exponential GARCH (REGARCH) models, and the quasi-maximum likelihood (QML) estimation. Commonly-used realized measures of return volatility and  their properties shall be discussed in Section 3. In Section 4, we shall discuss the volatility point forecast algorithm and forecast accuracy comparison framework, followed by an explanation of the data and a summary of the empirical results in section 5. Section 6 contains the conclusion and extension.

\section{Models and Estimation}

This section briefly illustrates the three conventional GARCH family models and realized GARCH-type models, while also explaining the terminology. In the GARCH-type models, the current volatility is represented as a function of the maginitude of the previous time periods' volatilities and error terms.
\begin{equation}
\begin{array}{lr}
	r_t=E(r_t|\mathcal{F}_{t-1})+ \epsilon_t\\
	\epsilon_t=\sqrt{h_t}z_t,\quad z_t\sim i.i.d.(0,1)
\end{array}
\end{equation}
 Daily return $r_t$ includes a predictable part $E(r_t|\mathcal{F}_{t-1})$ and an unpredictable part $\epsilon_t$, where $\{\mathcal{F}_t\}$ is a filtration, an information set available at time $t$. $z_t$ is an independently and identically distributed variable with mean of zero and standard deviation of one. In what follows, we set $E(r_t|\mathcal{F}_{t-1})=0$ because the null hypothesis of zero mean and that of no autocorrelation were not rejected in our data analysis. So that the conditional variance at time $t$ given previous information can be represented by $h_t=var(r_t|\mathcal{F}_{t-1})$, which is deterministic. On the contrary, $h_t^2$ is stochastic at time $t$ in the SV models, which is a critical difference between GARCH-type models and SV-type models.
\subsection{Conventional GARCH-type models }
The GARCH model suggested by Bollerslev (1986) relates the current volatility with the squares of previous innovations and previous conditional variances. The simplest GARCH model, the GARCH(1,1) model\nocite{GARCH}, is given by
\begin{equation}
h_t=\omega+\beta h_{t-1}+\alpha\epsilon_{t-1}^2\qquad\omega>0,\quad\alpha,\beta\geq0.
\end{equation}
Sign restrictions on $\omega$, $\alpha$, and $\beta$ ensure that $h_t$ is positive. The persistence on volatility is measured by $\beta$, and the previous innovation's persistence on volatility is measured by $\alpha$.

 However, the GARCH model is not sufficient to characterize stock market observations. Previous studies refer an asymmetry between equity volatility and returns, indicating that future volatility is usually higher when the market is performing poorly than when it is performing well. A competing explanation of volatility asymmetry is the so-called leverage effect, which dictates that a negative return increases financial and operating leverage, consequently increasing return volatility. Another class of explanation focuses on the volatility feedback effect. This effect states that if the market risk premium is an increasing function of market volatility, and if decreases in riskless rates do not offset increases in the market premia due to volatility increases, then increases in volatility should lead to drops in the market. Both GJR and EGARCH models capture the volatility asymmetry by using a leverage function. 

The EGARCH(1,1) model introduced by Nelson (1991)\nocite{EGARCH} is given by
\begin{equation}
\log h_t=\omega+\beta(\log h_{t-1}-\omega)+\tau(z_{t-1}).
\end{equation}
The leverage function, $\tau(\cdot)$, in the EGARCH model is defined as 
\begin{equation}
\tau(z_{t-1})=\tau_{1,1}z_{t-1,1}+\tau_{1,2}(|z_{t-1}|-E(|z_{t-1}|)).
\label{eq:leverage EGARCH}
\end{equation}
When $z_{t-1}\geq0$, equation (4) becomes
\begin{equation}
\log h_t=\omega-\tau_{1,2}E(|z_{t-1}|)+\beta(\log h_{t-1}-\omega)+(\tau_{1,1}+\tau_{1,2})z_{t-1}.
\end{equation}
When $z_{t-1}<0$, equation (4) becomes
\begin{equation}
	\log h_t=\omega-\tau_{1,2}E(|z_{t-1}|)+\beta(\log h_{t-1}-\omega)+(\tau_{1,1}-\tau_{1,2})z_{t-1}.
\end{equation}
If coefficient $\tau_{1,1}$ is negative,  $\tau_{1,1}z_{t-1}\leq0$ when $z_{t-1}\geq0$ and  $\tau_{1,1}z_{t-1}>0$ when $z_{t-1}<0$, which indicates that a negative return will increase the volatility the following day. Then we can infer that volatility asymmetry exists in the market. 

The GJR(1,1) \nocite{GJR}model by Glosten, Jagannathan and Runkle (1993) has the form:
\begin{equation}
h_t=\omega+\beta h_{t-1}+\alpha \epsilon_{t-1}^2+\tau(\epsilon_{t-1}),\quad\omega>0,\alpha,\beta\geq 0.
\label{eq:GJR}
\end{equation}
The leverage function in this model is defined as
\begin{equation}
\tau(\epsilon_{t-1})=\tau I_{\epsilon_{t-1}<0}\epsilon_{t-1}^2,\quad\tau\geq 0,
\label{eq:leverage GJR}
\end{equation}
where $I(\cdot)$ denotes the indicator function. When daily return $r_{t-1}$ is negative, $I=1$, and equation (\ref{eq:GJR}) becomes
\begin{equation}
	h_t=\omega+\beta h_{t-1}+(\alpha+\tau)\epsilon_{t-1}^2.
\end{equation}
When daily return $r_{t-1}$ is nonnegative, $I=0$, and equation (\ref{eq:GJR}) becomes
\begin{equation}
	h_t=\omega+\beta h_{t-1}+\alpha\epsilon_{t-1}^2.
\end{equation}
So, if $\tau>0$, a negative return is more prone than a positive return to cause high volatility the following day.

\subsection{Realized EGARCH models}
The REGARCH(1,1) model introduced by Hansen, Huang and Shek (2012)\nocite{RGARCH} is represented by:
\begin{equation}
\label{eq:RGARCH1}
\log h_t=\omega+\beta \log h_{t-1}+\gamma \log x_{t-1},
\end{equation}
\begin{equation}
\label{eq:RGARCH2}
\log x_t=\xi+\phi \log h_t+\tau(z_t)+u_t,
\end{equation}
where $x_t$ denotes the realized measure. Hansen et al. (2012) selected a quadratic specification leverage function, $\tau(z_t)=\tau_1 z_t+\tau_2(z_t^2-1)$, for its simplicity in quasi-likelihood analysis and the good statistical properties that allow this leverage function to ensure $E(\tau(z_t))=0$ as long as the distribution of $z_t$ satisfies with $E(z_t)=0$ and $Var(z_t)=1$. Equation (\ref{eq:RGARCH1}) is called the GARCH equation, while equation (\ref{eq:RGARCH2}) is called the measurement equation. The measurement equations , which relates the realized measure $x_t$ to the true variance $h_t$.

Hansen and Huang (2016)\nocite{JREGARCH} extended the realized GARCH model to a REGARCH model, which is more flexible in the modeling of the joint dependence of returns and volatility. The REGARCH model includes leverage functions in both of the GARCH  and measurement equations. The REGARCH model including K realized measures is given by the following equations:
\begin{equation}
\begin{array}{l}
r_t=\sqrt{h_t}z_t,\qquad z_t\sim N(0,1).\\
\log h_t=\omega+\beta( \log h_{t-1}-\omega)+\tau_1z_{t-1}+\tau_2(z_{t-1}^2-1)+\gamma u_{t-1},\\
\log x_{k,t}=\xi_k+\phi_{k}\log h_t+\delta_{k,1}z_t+\delta_{k,2}(z_t^2-1)+u_{k,t},\\
\mbox{where}\quad k=1,\dots,K.
\end{array}
\end{equation}
\begin{equation}
u_t=(u_{1,t},u_{2,t},\dots,u_{K,t})'\sim\mathcal{N}(0,\Sigma).
\end{equation}
\begin{equation}
\gamma=(\gamma_1,\dots,\gamma_K).
\end{equation}
The vector of realized measures, $x_{t}$, may include the squared return, RV, and RRV. $\gamma u_{t-1}$ is the main channel by which the realized measures drive the expectations of the logarithmic future volatility, $\log h_t$, up or down. Imposing the restrictions, $\phi_k=1$, $k=1,\dots,K$, makes it easier to interpret the model. $\xi_k$ is tied to the sampling error of the logarithmic realized measure, $\log x_k$. When we estimate a logarithmic daily close-to-close volatility, the logarithmic realized measure, $\log x_k$, that based on open hour high-frequency data is expected to be a downward biased measurement of $\log h_t$, $\xi_k$ is expected to be negative equivalently. A critical difference between the original RGARCH model and the REGARCH model is that the former one is limited to have the same coefficient for the leverage function, $\tau(z_t)$, and $u_t$ in the equation (12), while the later one is more flexible. 
\subsection{Estimation}
 Since the true data generation process of return is unknown (In Section 5.1, we reject the hypothesis that daily returns are distributed normally), the traditional maximum likelihood method, which assumes that the error term follows a specified density function (e.g., a normally distributed density function), may  result in likely specifcation errors. Thus, we estimate the above models using a QML method that maximizes the quasi-log-likelihood function. The quasi-maximum likelihood estimator (QMLE) of $\theta$ is the maximizer of the quasi-log-likelihood function. The QMLE is weakly consistent for the true parameter $\theta$.

For a simple case of two realized measures, we estimate the model underlying the assumptions of Gaussian specification:
\begin{itemize}
	\item $z_t\sim\mathcal{N}(0,1)$;
	\item  $u_t\sim\mathcal{N}(0,\Sigma)$; \quad
	$\Sigma=\left(\begin{array}{cc} \sigma_{11} & \sigma_{12} \\
	\sigma_{12}& \sigma_{22} \end{array} \right).$
\end{itemize}
\begin{equation}
r_t=\sqrt{h_t}z_t,\qquad z_t\sim N(0,1).
\end{equation}\begin{equation}
\log h_t=\omega+(\beta \log h_{t-1}-\omega)+\tau_1z_{t-1}+\tau_2(z_{t-1}^2-1)+\gamma u_{t-1}\quad\mbox{where}\quad \gamma=(\gamma_1,\gamma_2),
\label{eq:measurementeq}
\end{equation}
\begin{equation}
\log x_{i,t}=\xi_i+\phi_{i}\log h_t+\delta_{i,1}z_t+\delta_{i,2}(z_t^2-1)+u_{i,t}\quad \mbox{where}\quad i=1,2.
\label{eq:GARCHeq}
\end{equation}
By substituting the measurement equation (\ref{eq:measurementeq}) into the GARCH equation (\ref{eq:GARCHeq}), we can obtain 
\begin{equation}
\begin{array}{ll}
\log h_t=&\omega-\gamma_1\xi_1 -\gamma_2\xi_2+ (\beta-\gamma_1\phi_1-\gamma_2\phi_2)\log h_{t-1}\\
&+\gamma_1 \log x_{1,t-1}+\gamma_2 \log x_{2,t-1}+(\tau_1-\gamma_1\delta_{1,1}-\gamma_2\delta_{2,1})z_{t-1}\\
&+(
\tau_2-\gamma_1\delta_{1,2}-\gamma_2\delta_{2,2})(z_{t-1}^2-1).
\end{array}
\end{equation}
Therefore, the full-sample quasi-likelihood is calculated as follows:
\begin{equation}
\label{likelihood}
\begin{array}{rl}
\mathcal{L}(r,x;\theta,h_1)=&(\frac{1}{\sqrt{2\pi}})^{3T}(\frac{1}{|\Sigma|})^{T/2}\times\prod_{t=1}^{T}\frac{1}{\sqrt{h_t}}\times\\
&\prod_{t=1}^{T}[exp(-\dfrac{u_t'\Sigma
	^{-1}u_t}{2})exp(-\frac{r_t^2}{2h_t})].
\end{array}
\end{equation}

However, our primary interest lies in the log-likelihood for daily returns. By calculating the log-likelihood of returns, we can compare the estimation accuracy of the realized GARCH and REGARCH models against the conventional GARCH family models on the same scale.
\begin{equation}
l_{r}=-\frac{1}{2}\big(\frac{T}{\log{2\pi}}+\sum_{i=1}^{T}(\log(h_i)+\frac{r_i^2}{h_i})\big),
\end{equation}


\section{Realized Measures of Volatility}


This article focuses on two realized measures, RV (Andersen et al., 2001) and RRV (Christensen and Podolskij, 2007). This section briefly explains the relevant properties of these measures and methods.
\subsection{Basic Setup}
Assume that a logarithmic price $p$ along day t follows a continuous semimartingale sample path:
\begin{equation}
\label{eq:price}
p_t=p_0+\int_0^t\mu_udu+\int_{0}^{t}\sigma_udW_u\quad for\quad t\geq 0,
\end{equation}
 where $\mu$ is a predictable drift component, $\sigma$ is the instantaneous volatility, and $W$ is a standard Brownian motion. The correlation between $\sigma_t$ and $W_t$ indicates there is leverage effect.
The intraday return between sampling times $t_{i}$ and $t_{i+1}$, for  $0=t_0<t_1<\dots<t_n=1$ and $ i=0,...,n-1$, is denoted by $r_{t_i,\Delta_{i}}=p_{t_i}-p_{t_{i-1}}$, where $\Delta_{i}=t_i-t_{i-1}$.
The integrated variance (IV) is our object of interest because it is critcal to the financial economics of derivatives pricing and risk management. It is defined as 
\begin{equation}
IV=\int_0^1\sigma_u^2du.
\end{equation}
As IV is unobservable, substantial literature, including literature on RV and RRV, exists to measure IV.
\subsection{Realized Variance }

   The RV is defined as the sum of intraday high frequency squared returns, which is a non-parametric estimator of the IV.  RV builds on the idea of the quadratic variation of the logarithmic price process $p$, which is given by:
   \begin{equation}
   [p,p]=p\lim_{n\rightarrow\infty}\sum_{i=1}^n r_{t_i,\Delta_{i}}^2.
   \end{equation}
   In the framework of equation (\ref{eq:price}), the quadratic variation is coincident with  the IV. Therefore, under an equidistant partition over [0,1], $0=t_0<t_1<\dots<t_n=1$, where $t_i=i/n=i\Delta$,
\begin{equation}
	RV^{\Delta}=\sum_{i=1}^{n}r_{i\Delta,\Delta}^2.
\end{equation}
As $n\rightarrow\infty$, we have
\begin{equation}
RV\xrightarrow{p}IV.
\end{equation}

In Barndorff-Nielsen and Shephard (2002)\nocite{RV}, the distribution of RV is given by 
\begin{equation}
	\sqrt{n}(RV-IV)\xrightarrow{d}MN(0,2IQ),\quad IQ=\int_{0}^1\sigma_u^4du,
\end{equation}
where $MN$ denotes a mixed Gaussian distribution, and $IQ$ is the integrated quarticity.

\subsection{Realized Range-based Volatility}
Considering the microstructure noise effect, when $n$ is too large, the RV is both biased and inconsistent (Bandi and Russell, 2008; Hansen and Lunde, 2006). Previous studies have devoted considerable effort towards developing bias-reducing techniques with a focus on return-based realized measures. Aiming for higher precision than RV and using a different approach than the return-based estimations of IV, Christensen and Podolskij (2007) introduced RRV, which inspects all data points and prevents IV information neglect. In the framework of equation (\ref{eq:price}) and a presumption that $p$ is fully observed, the intraday range at sampling times  $t_{i-1}$ and $t_i$ is defined as follows:
\begin{equation}
S_{p_{t_{i}},\Delta_{i}}=\sup_{t_{i-1}\leq s,t\leq t_{i}}\{p_t-p_s\}.
\end{equation}
$S_{p_{t_{i}},\Delta_{i}}$ is also called the ``high-low" of $[t_{i-1},t_i]$.
The asymptomatic distribution of the range is derived using the theory of Brownian motion (Feller, 1951\nocite{Feller}) and the method deriving the moment-generating function of the range of Brownian motion (Parkinson, 1980\nocite{Parkinson}).
The range of a standard Brownian motion over $[t_{i-1},t_i]$ is defined as
\begin{equation}
S_{W_{t_{i}},\Delta_{i}}=\sup_{t_{i-1}\leq s,t\leq t_{i}}\{W_t-W_s\}.
\end{equation}
The theory of Feller and method of Parkinson enable the moment-generating function of the range of a scaled Brownian motion. For the $r$th moment of $S_{p_{t_{i}},\Delta_{i}}$, we have $\mathbb{E}[S_{p_{t_{i}},\Delta_{i}}^r]=\lambda_r\Delta_{i}^{r/2}\sigma^r,\quad r\geq1$, where $\lambda_r=\mathbb{E}[S^r_W]$.

The RRV is defined as
\begin{equation}
RRV^{\Delta}=\frac{1}{\lambda_{2}}\sum_{i=1}^{n}S^2_{pi\Delta,\Delta}.
\end{equation}

The distribution of the RRV is derived as
\begin{equation}
	Var(RRV)=\Lambda\sigma^4/n\quad \mbox{where }\Lambda=\frac{\lambda_{4}-\lambda_{2}^2}{\lambda_2^2}\approx 0.4073.
\end{equation}
When n is large enough, the probability limit of the RRV is also the IV.
\begin{equation}
RRV\xrightarrow{p}IV
\end{equation}
\begin{equation}
\sqrt{n}(RRV-\sigma^2)\xrightarrow{d}N(0,\Lambda\sigma^4)
\end{equation} 

The asymptotic variance of RRV (around 0.4IQ), is much smaller than that of RV (2IQ). By comparing the variance of RV and RRV for the same sampling frequency $n$, RRV is approximately five times as precise as RV, theoretically. We may assume, generally, that $mn+1$ equidistant observations of the price process are available, giving $mn$ returns. Thus, the observations are split into $n$ intervals, with each interval containing $m$ innovations.  The observed range of $p$ and $W$ over the $i$th interval are redefined as 
\begin{equation}
S_{p_{i\Delta,\Delta},m}=\max_{0\leq s,t\leq m}\{p_{(i-1)/n+t/mn}-p_{(i-1)/n+s/mn}\},
\end{equation}
\begin{equation}
S_{W,m}=\max_{0\leq s,t\leq m}\{W_{t/m}-W_{s/m}\}.
\end{equation}
The RRV is then generally noted by
 \begin{equation}
 RRV^{\Delta}_m=\frac{1}{\lambda_{2,m}}\sum_{i=1}^{n}S^2_{pi\Delta,\Delta,m}.
 \end{equation}

 Due to the tick-by-tick data of the Nikkei 225 stock index, the highest update frequency of price is 15s, which means the greatest value for $m$ is 20. $\lambda_{2,m}$ is a simulated constant that includes the downward bias to the true volatility caused by the limitation of $m$. One of the advantages of employing the REGARCH model with  RRV is that we can arbitrarily set an approximate value for $\lambda_{2,m}$ without a simulation. The bias caused by $\lambda_{2,m}$ can be justified by $\xi$ in the measure equation.\\





\section{Volatility Forecasting}
In the GARCH model, volatility $\sigma_{t}$ is known at time $t-1$. The REGARCH model inherits this feature, making it easy to calculate the one-day-ahead forecast volatility.  In this section, we will illustrate the one-step-ahead point-forecast schemes and the method for forecasting accuracy comparison. We will adopt a recursive and a rolling forecasting scheme.

In the recursive forecasting scheme, by setting the beginning point of sampling to be $k$, the one-step-ahead forecast of log conditional variance is calculated as follows:
\begin{itemize}
	\item \textit{Step 1.}  Set $i$ = 1.
	\item\textit{Step 2.} Estimate the parameters of the REGARCH model using the sample\\
	 $\{r_{1},\dots,r_{i+k-1},x_{j,1},\dots,x_{j,i+k-1}\}$ by QML ($j$ is the type of $x$).
	\item\textit{Step 3.} Set the parameters $\mathbf{\theta}=\{\omega,\beta,\tau,\gamma,\xi,\delta\}$ in the REGARCH model equal to their estimates obtained in \textit{step 2.} Then calculate $\hat{\log h}_{k+i}$ by substituting $\hat{\log h}_{k+i-1}$ and $\log x_{j,k+i-1}$.
	\begin{equation}
	\begin{array}{rl}
		\hat{\log h}_{k+i}=&(1-\hat{\beta})\hat{\omega}	+\sum_{j=1}^K\{-\hat{\gamma}_j\hat{\xi}_j+\hat{\gamma}_j*\log x_{j,k+i-1}+(\hat{\beta}-\hat{\gamma}_j)\hat{\log h}_{k+i-1}	\\
	&+(\hat{\tau}_1-\hat{\gamma}_j\hat{\delta}_{j,1})\hat{z}_{k+i-1}+(\hat{\tau}_2-\gamma_j\hat{\delta}_{j,2})(\hat{z}_{k+i-1}^2-1)\},
	\end{array}
	\end{equation}
	where $\hat{z}_{k+i-1}=\dfrac{r_{k+i-1}}{\exp(\hat{\log h}_{k+i-1}/2)}$.
	\item\textit{Step 4.} Set $i=i+1$ and return to \textit{Step 1.} if $i<T-k$, ending if $i=T-k$.
\end{itemize}
Using the above steps allows the generation of the one-step-ahead point-forecast. \\

In rolling window forecasting, models are reestimated daily by recursively adding one observation to the end of the sample and removing one from the start. The width of the rolling window is fixed as $K$. We can obtain the point-forecast value of the logarithmic daily return volatility as follows:

\begin{itemize}
	\item \textit{Step 1.}  Set $i$ = 1.
	\item\textit{Step 2.} Estimate the parameters of the REGARCH model using the sample\\ $\{r_{1+i},\dots,r_{i+k-1},x_{j,1+i},\dots,x_{j,i+k-1}\}$ by QML ($j$ is the type of $x$).
	\item\textit{Step 3.} Set the parameters $\mathbf{\theta}=\{\omega,\beta,\tau,\gamma,\xi,\delta\}$ in the REGARCH model equal to their estimates obtained in \textit{step 2.} Then calculate $\hat{\log h}_{k+i}$ by substituting $\hat{\log h}_{k+i-1}$ and $\log x_{j,k+i-1}$.
	\begin{equation}
	\begin{array}{rl}
	\hat{\log h}_{k+i}=&(1-\hat{\beta})\hat{\omega}	+\sum_{j=1}^K\{-\hat{\gamma}_j\hat{\xi}_j+\hat{\gamma}_j*\log x_{j,k+i-1}+(\hat{\beta}-\hat{\gamma}_j)\hat{\log h}_{k+i-1}	\\
	&+(\hat{\tau}_1-\hat{\gamma}_j\hat{\delta}_{j,1})\hat{z}_{k+i-1}+(\hat{\tau}_2-\gamma_j\hat{\delta}_{j,2})(\hat{z}_{k+i-1}^2-1)\},
	\end{array}
	\end{equation}
	where $\hat{z}_{k+i-1}=\dfrac{r_{k+i-1}}{\exp(\hat{\log h}_{k+i-1}/2)}$.
	\item\textit{Step 4.} Set $i=i+1$ and return to \textit{Step 1.} if $i<T-k$, ending if $i=T-k$.
\end{itemize}

\section{Empirical Results }
\subsection{Data Description}

We used daily data on returns and high-frequency data on realized volatilities for the Nikkei 225 stock index. The sample period ran from January 4, 2010, to July 14, 2017, delivering 1848 days (T=1848). The data were obtained from the Nikkei Economic Electronic Databank System Tick-Data. The plain realized variance is calculated as the sum of the squared intraday returns using a five-minute frequency. The RRV was calculated as the sum of the squared difference between five-minute low and high prices using the method proposed by Christensen and Podolskij (2007)\nocite{RRV}. The daily return was calculated as a close-to-close return. In the volatility forecasting, we set $K$ as 1386 (the window width in the rolling-window or the start point in the recursive forecast), three-quarters of the full sample size, setting 548 as the out-of-sample size.

In empirical work, for realized measures, the benefits of more frequent sampling are traded off against the cumulated noise. Liu et al. (2015)\nocite{5min} proposed realized measures based on frequencies of five minutes led to increased accuracy. Therefore, we calculated the realized measures in five-minute frequencies. It should be noted that the calculation of RRV is simplfied here. As per the explanation in Section 3, we calculated the RRV as follows:
\begin{equation}
RRV_t=1/\lambda_{2,m}\sum_{i=0}^n(\log(high_{t-1+(i+1)/n})-\log(low_{t-1+i/n}))^2.
\end{equation}
To obtain the value of $\lambda_{2,m}$, as illustrated in Christensen and Podolskij (2007), a simulation is required. However, an arbitrary approximate value was chosen for this article since the bias can be absorbed by $\xi$, the constant term in the measurement equation, giving $\lambda_{2,m}=2$.

\begin{figure}[!]
	\centering
	\includegraphics[width=\linewidth]{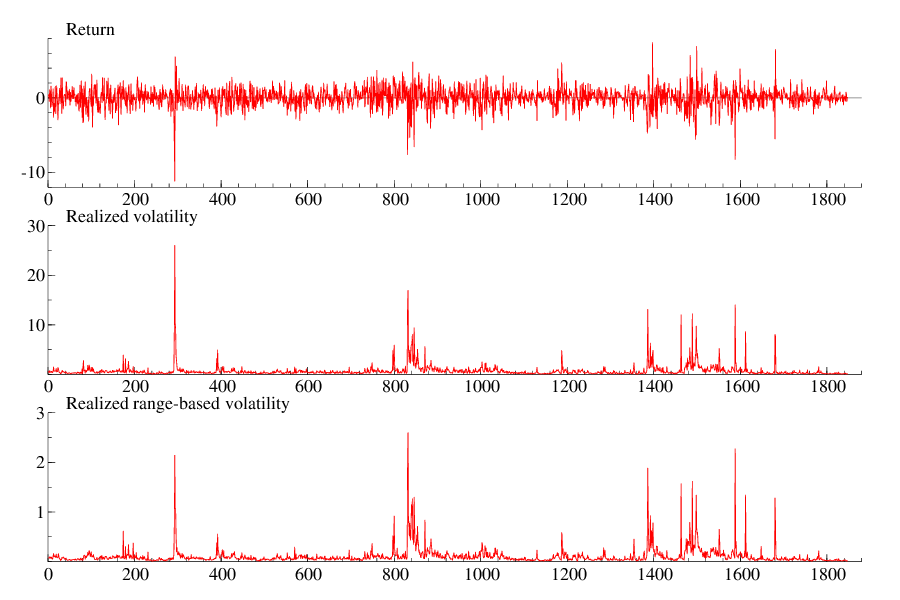}
	\caption{Daily returns, logarithmic-realized volatility, and logarithmic-realized  range-based volatility of the Nikkei 225 stock index}
	\label{fig:figvolrv5rrv5}
\end{figure}

\begin{figure}[!]
	\begin{center}
		\includegraphics[width=\linewidth]{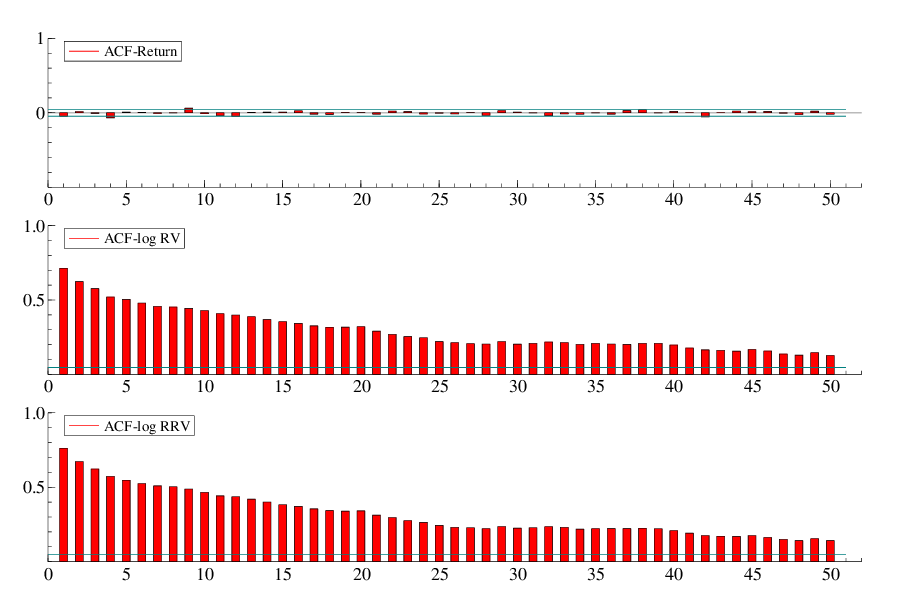}
		\caption{Autocorrelation}
		\label{fig:figAFC}
	\end{center}
\end{figure}

From Figure \ref{fig:figvolrv5rrv5}, we can observe that the log-realized volatility and log-realized range-based volatility acted coincidently most of the time, yet acted differently when the jumps occurred. Table \ref{tab:Descriptive}  shows the descriptive statistics of the daily returns in percentage, log-realized variance, and log-realized range-based volatility.  The daily return skewness is significantly negative, indicating an asymmetric tail extending toward negative return. The daily return and realized measures kurtoses are significantly greater than three, which means the distribution of $z$ is more peaked than a normal distribution. The kurtosis statistic confirms the well-known phenomenon that the distribution of the daily return is leptokurtic, possessing ``fat tails". We can assume a skewed t-distribution for returns, which we have left for future research.  The Jarque-Bera (JB) statistic, which uses both skewness and kurtosis, also rejects the null hypothesis of normality at the 1\% significance level. LB(10) is the Ljung-Box statistic, adjusted for heteroskedasticity to test the null hypothesis of no autocorrelations up to 10 lags. According to the daily return LB statistic, the null hypothesis cannot be rejected at the 10\% significance level ($21.57<23.209$).  The log-realized variance and log-realized range-based volatility are strongly rejected at the $1\%$ significant level. This strongly auto-correlated characteristic is shown more visually in Figure \ref{fig:figAFC}.


\begin{table}[!]
		\caption[Descreptive]{Descriptive Statistics of the Nikkei 225 stock index}
	\resizebox{\textwidth}{!}{	\begin{tabular} {l cccccccc}
			\toprule
			&          Mean     &     SD    &     Max  &     Min       &    Sk        &    Ku     &  JB    &  LB(10)\\\midrule
			 daily return (\%)                           &          0.034    &   1.392   &   7.426   &     -11.153 & -0.540    &    8.226     &   2192.91    &   21.57\\
			&          (0.032) &             &           &            &   (0.057)      &     (0.114)     &           &        \\
			log-realized volatility          &       -0.779   &  0.841 & 3.258&   -3.189  &   0.738  & 4.470 &  334.02 &  5163.52\\
			&         (0.020) &            &           &            &   (0.057)      &     (0.114)      &             &         \\ 
			log-realized range-based volatility  &       -2.629    & 0.804 &0.955 & -4.817    & 0.810   &4.482  & 371.15 &  6107.43 \\
			&          (0.019) &           &           &            &   (0.057)      &     (0.114)        &             &         \\		
			\bottomrule
	\end{tabular}}
	\label{tab:Descriptive}
\end{table}

\subsection{Goodness of Fit}


\begin{table}[!]
			\caption{Estimation results}
		\resizebox{\textwidth}{!}{\begin{tabular}{l ccccc ccccc ccccc }
			\toprule
			& $ \omega $ &	    $  \beta $    &	  $ \tau_1 $ ($\alpha$)   &	$   \tau_2$ ($\tau$) &	  $  \gamma_1$    &$	  \gamma_2  $  &	$     \xi1$&	$	\xi2 $ &	  $\delta_{1,1}$&	  $  \delta_{1,2} $  &	$ \delta_{2,1}  $ &	 $\delta_{2,2}$ &	   $\sigma_{11}$  &	  $\sigma_{22}$&		$\sigma_{12}$\\\hline\\
		GARCH&0.077 &   0.830    &   0.137&&&&&&&&&&&&\\
		&(0.025)    &    (0.028)   &     (0.025)&&&&&&&&&&&&\\\\
			
			GJR&0.087  &  0.833    &   0.055   &    0.143&&&&&&&&&&&\\
		&(0.027)     &  (0.024)    &   (0.031)   &    (0.054)&&&&&&&&&&&\\\\
			
			EGARCH&	      0.645  &     0.941   &  -0.113   &    0.227&&& &&&&&&&\\	
	&      (0.133)    &   (0.016) &      (0.036) &      (0.039)&&&&& &&&&&\\\\

		REGARCH-$r^2$ &	       0.466    &   0.954   &   -0.033&       0.094    &   0.036&   &  -1.517 &   &   0.244  &       0.686 &&     & 3.939&&\\
		&     (0.136)  &     (0.016)     &  (0.044)  &     (0.013)     &  (0.012) & &   (0.052)& &  (0.047)  &  (0.070)  & & &   (0.281)&&\\\\


			
			REGARCH-RV5  &         0.477     &  0.924     & -0.114  &     0.062  &     0.313&   &  -1.228  &&   -0.127       & 0.098  &   &&  0.232&&\\		
			&    (0.098)    &   (0.014)     &  (0.012)    &   (0.010)  &     (0.033)  &  &   (0.046)     &  &(0.012)   &    (0.009)  &  &&   (0.011)\\\\		
			
		

			REGARCH-RRV5  & 0.477   &     0.925   &    -0.117  &     0.062  &      0.363   &&    -3.078 &&      -0.121    &     0.088   &   & &   0.183& & \\
			
		  &  (0.097)  &      (0.012)   &     (0.011)   &     (0.010)   &  (0.033)   &  &      (0.045)   & &    (0.011)  &     (0.008)     &  &&(0.010)\\\\

			 	JREGARCH-RV5,RRV5&        0.477   &       0.927     &    -0.116  &        0.058    &     -0.082  &        0.453     &    -1.229     &     -3.079      &   -0.126    &      0.102  &       -0.120   &       0.088    &      0.234    &      0.183   &   0.193\\
			 		
			 		     &    (0.097)     &     (0.011)       &   (0.011)    &      (0.010)    &      (0.049)   &       (0.063)   &       (0.045)   &    (0.045)   &     (0.012)    &     (0.009)   &       (0.011)    &      (0.008)       &   (0.011)     &     (0.010)   &     (0.010)\\\\
			 		                      
			 	
	\bottomrule			\end{tabular}}
\label{tab:Estimate_J}	
\begin{footnotesize}
Notes: Sample period: 4 January 2010–14 July 2017. Sample size: 1848. The numbers in parentheses are QML standard errors.
\end{footnotesize}
\end{table}

Table \ref{tab:Estimate_J} reports the estimation results for three GARCH-type models and the REGARCH models. The numbers in parentheses are QML standard errors. We estimated the REGARCH models with and without the condition $\phi_k=1$, $k=1,2$. However, since each $\phi_k$ was not significantly divergent from $1$, we adopted the REGARCH including the condition of $\phi_k=1$. The significance of the coefficient, $\gamma_k$, revealed the gain from including $u_{t-1}$ in to equation (13), to extract volatility-related information. It denoted that current volatility was strongly related to the lagged realized measure, an equivalent to the lagged measurement error. In a REGARCH model, since the information of past return has already been contained in the GARCH equation, we should not expect benefit from including the squared past daily return, $r^2_t$, as a measure of volatility in the measurement equation. So, in the REGARCH-$r^2$ model, the coefficient, $\gamma$, is extremely small. In addition, the model-specific parameters, $\tau_{1}$ and $\delta_{k,1}$, were both positive and strongly significant, indicating that $x_{k,t}$ and $h_{t+1}$ will be larger when $z_t<0$ than when $z_t>0$. This fact confirmed the stylized ``leverage effect" of the stock market. The negative coefficient $\xi_k$ can be explained as a downward bias between realized measures and true volatility. Since we used close-to-close daily returns and realized measures calculated using the intraday returns only when the market is open, the realized measures undervalued the volatility.  Moreover, we can compare the efficiency of the RV and RRV as a volatility measure by comparing $\sigma_{11}$ and $\sigma_{22}$ in the joint realized EGARCH model. In our analysis, the measurement error of the RRV, $\sigma_{22}$, is appreciably smaller than the measurement error of the RV, $\sigma_{11}$. We calculated the correlation coefficient between the RV and RRV as $0.934$, indicating that the RRV and RV are correlated, but contain different dynamics. In the REGARCH-$r^2$ model, the coefficient, $\sigma_{11}$, was considerably large, which showed that the squared past daily return was a ``bad" measure for volatility.

To evaluate the sufficiency of our proposed models, we used the Akaike's Information Criterion (AIC), Schwarz's Bayesian Information Criterion (SBIC), and the full-sample return log-likelihood function.
The criteria are calculated as follows:
\begin{equation}
\begin{array}{c}
	AIC=-2\ln L + 2n,\\
SBIC=-2\ln L+n\ln(T).
\end{array}
\end{equation}
 where $n$ is the number of estimated parameters. The AIC, SBIC, and return log-likelihood values of conventional GARCH-type and REGARCH-type models are reported in Table \ref{tab:goodness}. The return log-likelihood values showed that REGARCH-type models improved the goodness of fit. According to all criteria, the use of RRV as the realized measure was preferable in the REGARCH model specification. Due to the AIC and SBIC, the fit goodness of a joint REGARCH was superior to the fit goodness of an REGARCH, which included only one realized measure. Moreover, it was convinced again that the squared past daily return was a poor proxy for volatility when compared to high-frequency realized measures. 
\begin{center}	
	\begin{table}[!]
		\centering
		\caption{Goodness of fit}
		\begin{tabular}{l ccc}
			\toprule
			Model   &AIC&SBIC&$l_r$\\\midrule	
			GARCH&&&-3081.99\\
			EGARCH&  &  &  -3054.19\\
			GJR&&&-3063.71\\
					REGARCH-$r^2$& 13918.91&     13968.61& -3061.51\\	
			REGARCH-RV5& 8642.54&    8692.23  &-3039.70\\
			REGARCH-RRV5& 8197.52&  8247.22& -3038.98\\	
			JREGARCH-RV5,RRV5&7006.75&7089.58&-3040.01\\\bottomrule
		\end{tabular}
		\label{tab:goodness}
	\end{table}
\end{center}
\subsection{Forecast Evaluation}

Theoretically, a creditworthy forecasting model should not only fit the data
well, but, more importantly, should also
yield superior forecasting performance in the out-of-sample period. However, several difficulties were faced when comparing forecast accuracy. First, true volatility is unobservable, and the loss functions are computed using an imperfect volatility proxy. Using a noisy volatility proxy in some commonly-used tests for forecast comparison can give rise to distortions in the ranking of competing forecasts. Second, extreme observations may have a considerable impact on the outcomes of forecast evaluation and comparison. In Patton (2011), two robust forecast loss functions were introduced to solve this problem: mean squared error (MSE) and quasi-likelihood (QLIKE). As long as the volatility proxy is a conditionally unbiased volatility estimator, those two loss functions can return a consistent ranking of competing models. For the second problem, it should be noted that the MSE is prone to influence from extreme observations and the level of volatility that gauges the forecast accuracy (Patton and Sheppard, 2009; Patton, 2011).\nocite{Patton2009} Previous studies (Hansen and Lunde, 2005; Patton and Sheppard, 2009; Patton, 2011) have suggested that using QLIKE increases the power to reject inferior estimators. In the remainder of this subsection, we shall evaluate the out-of-sample forecasting ability of the REGARCH models against MSE, QLIKE, and out-of-sample return log-likelihood ($l_{r, out}$) criteria. We calculate those three loss functions as follows:\\
 \begin{equation}
 \label{eq:MSE}
 MSE= (T-k)^{-1}\sum_{t=k+1}^{T}(\hat{h}_t-\hat{\sigma}_t^2)^2
 \end{equation}
 
 \begin{equation}
 \label{eq:QLIKE}
 QLIKE= (T-k)^{-1}\sum_{t=k+1}^{T}\Big(\frac{\hat{\sigma}_t^2}{\hat{h}_t^2}-\log \frac{\hat{\sigma}^2_t}{\hat{h}_t}-1\Big)
 \end{equation}
 
 \begin{equation}
 l_{r,out}=-\frac{T-k}{2\log{2\pi}}-\frac{1}{2}\sum_{j=k+1}^{T}\Big(\log(\hat{h}_j)+\frac{r_i^2}{\hat{h}_j}\Big)
 \end{equation}

We used the RK calculated by the method of Barndorff et al. (2011)\nocite{RK2011}, as the proxy of volatility. As the RK has taken microstructure noise into account, we calculate the RK using one-minute returns to obtain as much information as possible. To adjust the bias caused by non-trading hours in the RK, we adopted the method proposed by Hansen and Lunde (2005)\nocite{c} to calculate unbiased proxy:
 \begin{equation}
 \hat{\sigma}^2_t\equiv\hat{c}RK_t^o,
 \end{equation}  
 \begin{equation}
  \hat{c}=\frac{\sum_{t=1}^{n}(r_t-\bar{r}_t)^2}{\sum_{t=1}^{n}RK_t^o}.
 \end{equation}
 where $RK_t^o$ is the RK that can ignore non-trading hours.
 
Panel A of Table \ref{tab:forecast evaluation}. shows the forecast results under the recursive scheme, and panel B shows the results under the rolling-window scheme with a fixed length of 1386 sample days. There are several salient features. First, the REGARCH models including high-frequency realized measures outperformed the conventional GARCH family models in the QLIKE and $l_{r, out}$ criteria, while the MSE of the EGARCH model was exceptional. The MSE depended on $\hat{h}_t-\hat{\sigma}_t^2$, which is centered around zero but has a proportional variance to the square of the variance of returns. The contradictive MSE value of the EGARCH model can be explained as being sensitive to extreme volatility jumps and spikes (see the red lines in Figures 3 and 4 ), which is coincident with previous research providing further motivation for the use of QLIKE as opposed to MSE in volatility forecasting applications.
 Second, the REGARCH models including RRV outperformed the REGARCH models including RV for all statistical criteria. This is consistent with the findings from the full-sample estimation where the RRV was more efficient than RV. Third, the joint REGARCH model dominated in the QLIKE and $l_{r, out}$ criteria provided strong evidence that the inclusion of multiple realized measures enhances the forecast quality. All of the findings have been proved robust against both the recursive and rolling-window forecast.

\begin{table}[!]
	\centering
	\caption{Forecast evaluation}
	\begin{subtable}[t]{\linewidth}
		\centering
		\vspace{0pt}
		\begin{tabular}{@{\extracolsep{5pt}} lccc}
			\toprule
			models& MSE & QLIKE &  $l_{r,out}$\\
			\midrule
			GARCH & $13.0002$ & $0.3409$ & $-784.7335$  \\
				GJR&$12.4101$ & $0.3195$ & $-774.5665$  \\
			EGARCH & $11.7131$ & $0.3031$ & $-770.0537$  \\
				REGARCH-$r^2$& $13.1920$ & $0.3174$ & $-779.9357$  \\
		REGARCH-RV5& $14.0892$ & $0.2898$ & $-767.3170$  \\
		REGARCH-RRV5 & $14.3181$ & $ 0.2791$ & $-765.4164$  \\
		JREGARCH-RV5,RRV5 & $13.4921$ & $0.2785$ & $-765.1140$\\
			\bottomrule
		\end{tabular}
		\caption{Panel A: recursive forecast}
	\end{subtable}
	\begin{subtable}[t]{\linewidth}
		\centering
\begin{tabular}{@{\extracolsep{5pt}} lccc}
	\toprule
	models& MSE & QLIKE &  $l_{r,out}$\\
	\midrule
	GARCH & $13.0482$ & $0.3436$ & $-785.6757$  \\
	GJR&$12.6235$ & $0.3234$ & $-775.7198$  \\
	EGARCH & $11.8486$ & $0.3059$ & $-771.2465$  \\
	REGARCH-$r^2$& $13.2238$ & $0.3175$ & $-780.0811$  \\
	REGARCH-RV5& $ 16.3475 $ & $0.2905$ & $ -767.5624$  \\
	REGARCH-RRV5 & $15.5979$ & $0.2799$ & $-765.6494$  \\
	JREGARCH-RV5,RRV5 & $14.7067 $ & $0.2775$ & $-765.0600$\\
	\bottomrule
\end{tabular}
		\caption{Panel B: rolling window forecast}
	\end{subtable}
\label{tab:forecast evaluation}
\end{table}

\begin{figure}[!]
		\begin{multicols}{2}
	\begin{subfigure}[t]{0.5\textwidth}
		\centering
		\includegraphics[width=\textwidth]{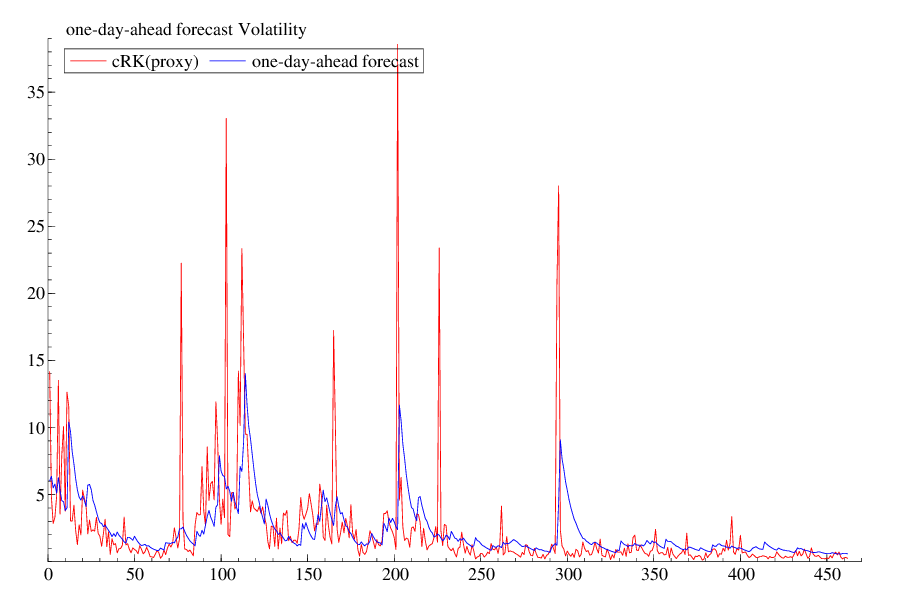} 
		\caption{GARCH} \label{fig:foreGARCH}
	\end{subfigure}		
	\begin{subfigure}[t]{0.5\textwidth}
		\centering
		\includegraphics[width=\textwidth]{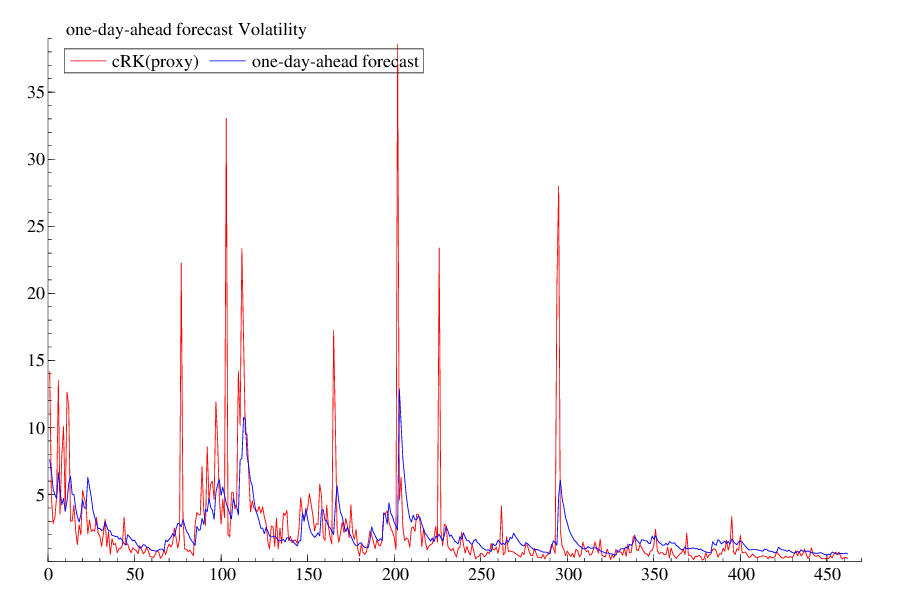} 
		\caption{EGARCH} \label{fig:foreEGARCH}
	\end{subfigure}
\end{multicols}

	\begin{multicols}{2}
		\begin{subfigure}[t]{0.5\textwidth}
		\centering
		\includegraphics[width=\textwidth]{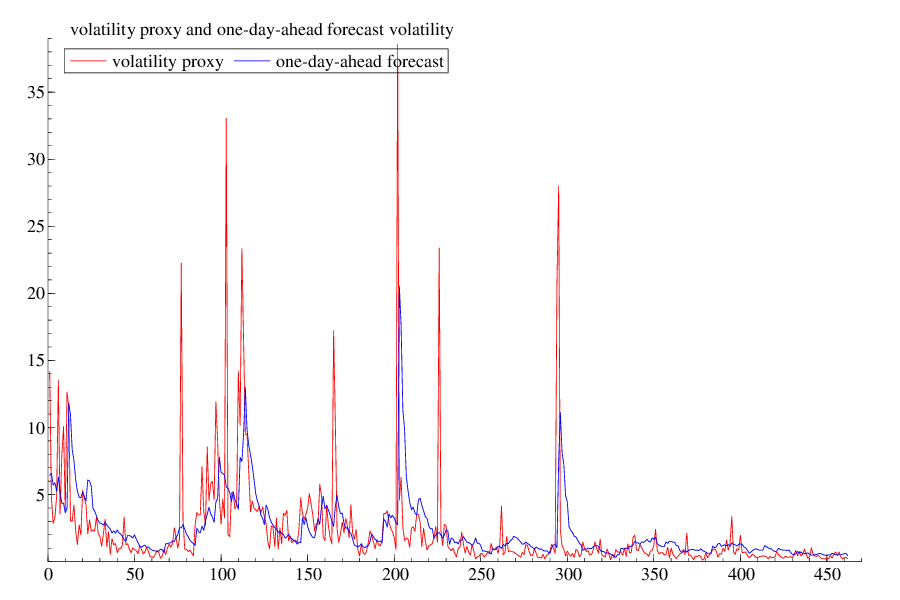} 
		\caption{REGARCH-$r^2$} \label{fig:forer^2}
	\end{subfigure}
	
	\begin{subfigure}[t]{0.5\textwidth}
		\centering
		\includegraphics[width=\textwidth]{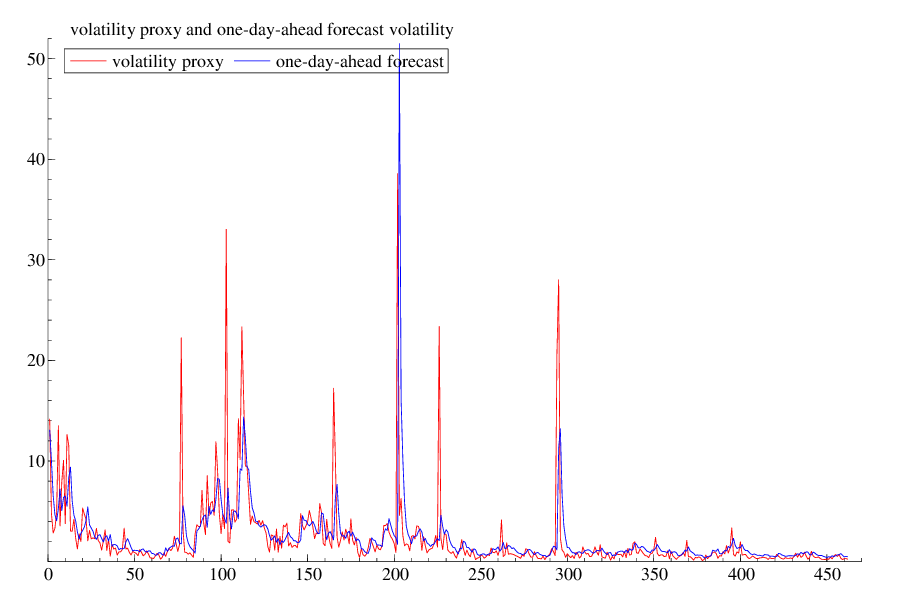} 
		\caption{REGARCH-RV 5min} \label{fig:foreRV}
	\end{subfigure}
	\end{multicols}
	\begin{multicols}{2}
	\begin{subfigure}[t]{0.5\textwidth}
		\centering
		\includegraphics[width=\textwidth]{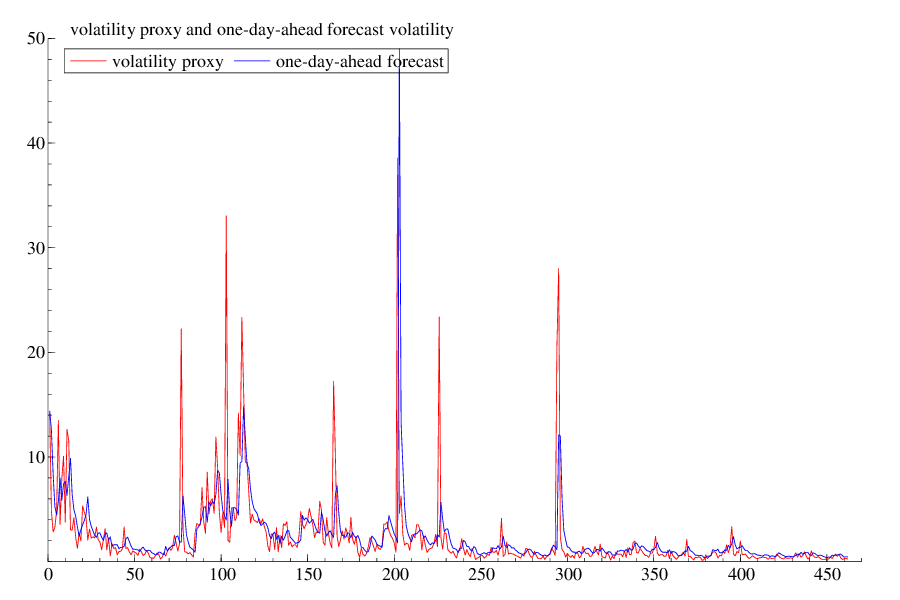} 
		\caption{REGARCH-RRV 5min} \label{fig:foreRRV}
	\end{subfigure}
	
	\begin{subfigure}[t]{0.5\textwidth}
		\centering
		\includegraphics[width=\textwidth]{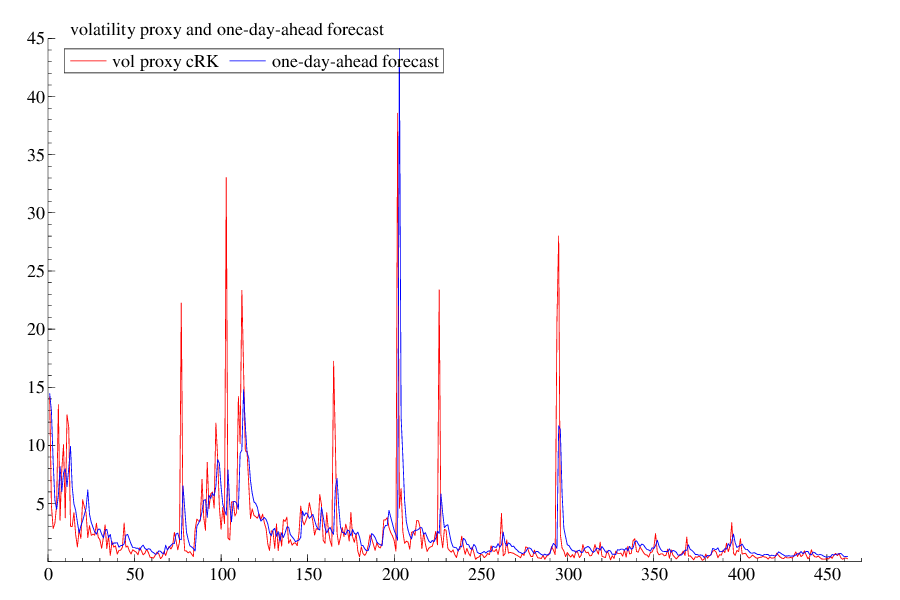} 
		\caption{JREGARCH-RV 5min, RRV 5min } \label{fig:foreRVRRV}
	\end{subfigure}
\end{multicols}
	\caption{Rolling-window one-day-ahead volatility forecast}
	\begin{footnotesize}
				Note: The blue lines denote one-day-ahead volatility forecasts and the res lines denote the volatility proxy.
	\end{footnotesize}
	\label{fig:rollingwindow}
\end{figure}

\begin{figure}[!]
	\begin{multicols}{2}
		\begin{subfigure}[t]{0.5\textwidth}
			\centering
			\includegraphics[width=\textwidth]{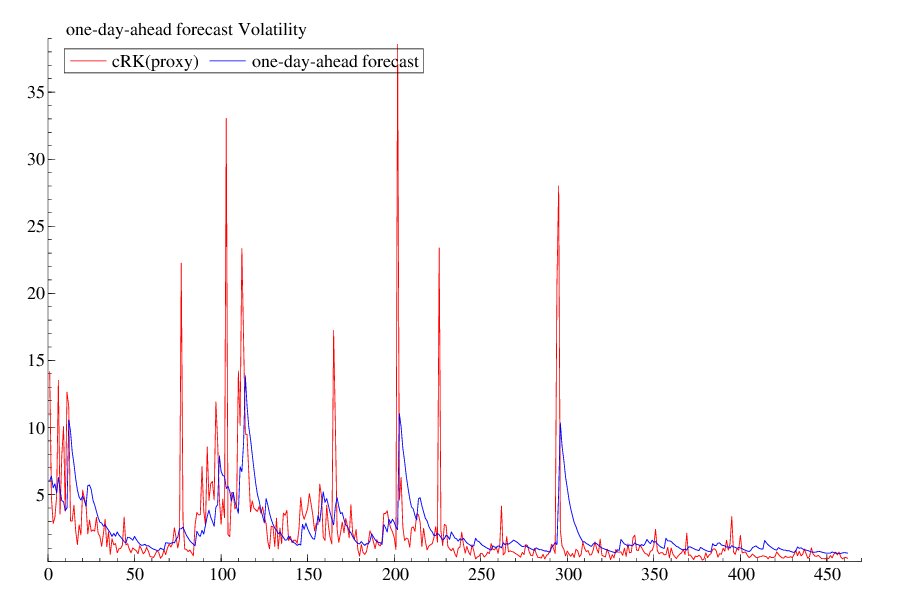} 
			\caption{GARCH} \label{fig:foreGARCHr}
		\end{subfigure}		
		\begin{subfigure}[t]{0.5\textwidth}
			\centering
			\includegraphics[width=\textwidth]{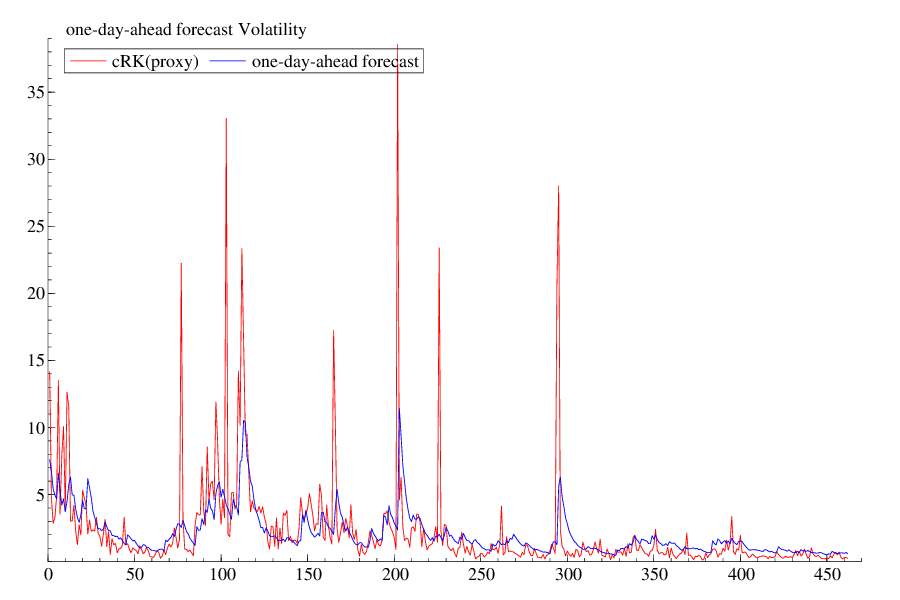} 
			\caption{EGARCH} \label{fig:foreEGARCHr}
		\end{subfigure}
	\end{multicols}
	
	\begin{multicols}{2}
		\begin{subfigure}[t]{0.5\textwidth}
			\centering
			\includegraphics[width=\textwidth]{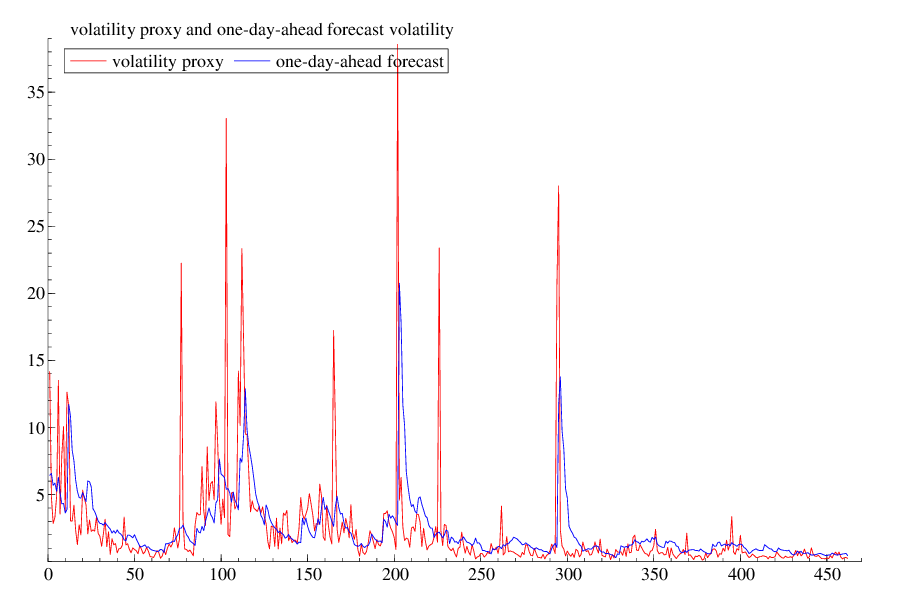} 
			\caption{REGARCH-$r^2$} \label{fig:forer^2r}
		\end{subfigure}
		
		\begin{subfigure}[t]{0.5\textwidth}
			\centering
			\includegraphics[width=\textwidth]{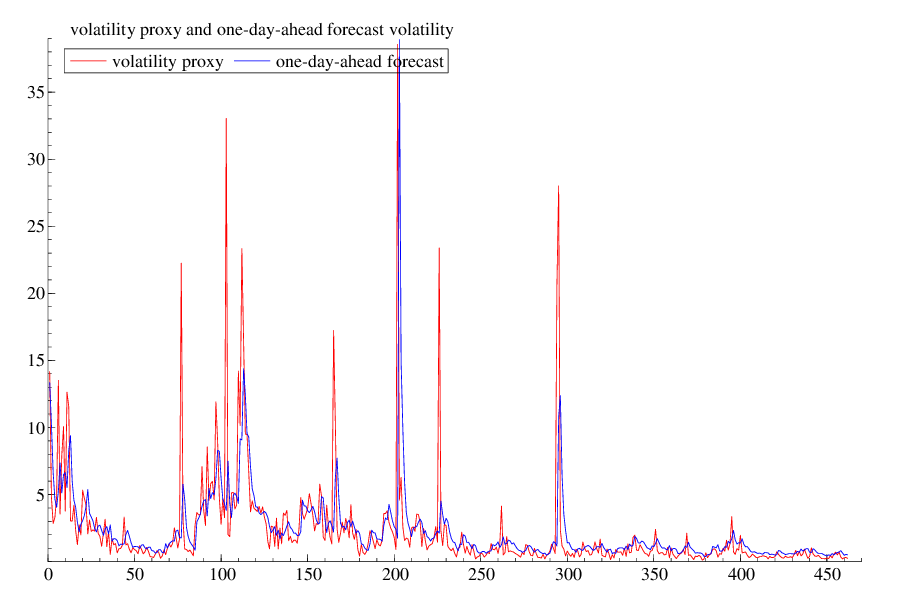} 
			\caption{REGARCH-RV 5min} \label{fig:foreRVr}
		\end{subfigure}
	\end{multicols}
	\begin{multicols}{2}
		\begin{subfigure}[t]{0.5\textwidth}
			\centering
			\includegraphics[width=\textwidth]{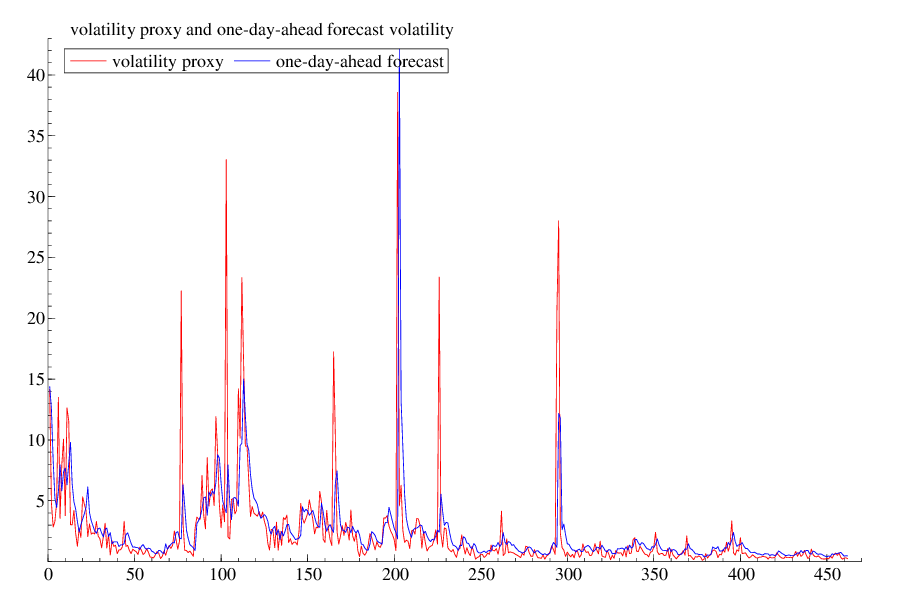} 
			\caption{REGARCH-RRV 5min} \label{fig:foreRRVr}
		\end{subfigure}
		
		\begin{subfigure}[t]{0.5\textwidth}
			\centering
			\includegraphics[width=\textwidth]{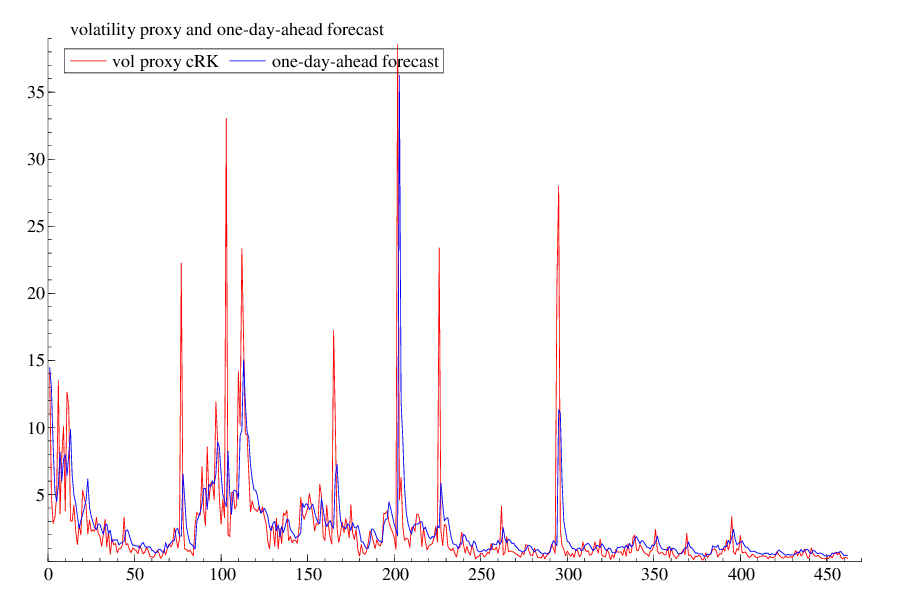} 
			\caption{JREGARCH-RV 5min, RRV 5min } \label{fig:foreRVRRVr}
		\end{subfigure}
	\end{multicols}
	\caption{Recursive forecast one-day-ahead volatility forecast}
	\label{fig:recursive}
	\begin{footnotesize}
		Note: The blue lines denote one-day-ahead volatility forecasts and the res lines denote the volatility proxy.
	\end{footnotesize}
\end{figure}

By comparing Figures \ref{fig:rollingwindow} and \ref{fig:recursive}, it is observable that the forecast volatility of the REGARCH models with high-frequency data-based volatility measures outperformed GARCH-type models, represented by the more volatile blue lines. Comparatively, the rolling-window scheme was prone to overvalue the volatility at extremely volatile moments, while the recursive scheme alleviated this overestimation.

\section{Conclusion and Extension}
Our empirical results demonstrate that incorporating incremental dynamic volatility information from realized measures improves both in-sample model fit and out-of-sample forecast accuracy. Specifically, we provide evidence that setting $\phi=1$ enhances the performance of REGARCH models, and that RRV is a more effective measure than RV. Models incorporating RRV yield superior out-of-sample forecasting accuracy compared to those using RV, with robustness confirmed under both rolling-window and recursive evaluation schemes. Overall, this study contributes to volatility modeling and forecasting by highlighting the advantages of employing a joint REGARCH framework for one-step-ahead volatility predictions.
Notably, the realized GARCH-type models used in this study are also well-suited for multi-step-ahead forecasting, provided that the dynamic properties of the realized measures are appropriately specified. Future research could extend this work by exploring volatility predictions over longer horizons and incorporating jump-robust estimators to enhance forecasting precision.


\clearpage

\end{spacing}
\end{document}